\documentclass[manuscript]{aastex63}

\usepackage{upgreek}
\usepackage{graphicx}
\usepackage{natbib}
\usepackage{amsmath}
\usepackage[version=3]{mhchem} 
\usepackage{hyperref}

\newcommand{\upd}{\mathrm{d}}
\newcommand{\2}{$_2$}
\newcommand{\pa}{$p_{a0}$}

\shorttitle{}

\begin{document}

\title{The phase-curve signature of condensible water-rich atmospheres on slowly-rotating tidally-locked exoplanets}


\correspondingauthor{Feng Ding}
\email{fengding@g.harvard.edu}

\author[0000-0001-7758-4110]{Feng Ding}
\affiliation{School of Engineering and Applied Sciences, Harvard University, Cambridge, MA 02138, USA}

\author[0000-0002-5887-1197]{Raymond T. Pierrehumbert}
\affil{Department of Physics, University of Oxford, Oxford OX1 3PU, UK }



\begin{abstract} 

We use an idealized three-dimensional general circulation model to study condensible-rich atmospheres with an ineffective cold trap on slowly-rotating tidally-locked terrestrial planets. In particular, we show the climate dynamics in a thin and temperate atmosphere with condensible water vapor. The similarities between our thin and temperate atmosphere and the warm and thick atmosphere approaching the water vapor runaway greenhouse in previous work are discussed, including the reversal of the thermal emission between the day and night hemisphere. Different from the transit spectroscopy of water vapor that depends on the absolute amount of atmospheric water vapor, the contrast between the dayside and nightside thermal emission provides information regarding the relative ratio of water vapor to the background atmosphere as well as the atmospheric pressure near the substellar tropopause and the emission level on the nightside on potentially habitable worlds. 


\end{abstract}

%
%

\keywords{astrobiology --- methods: numerical --- techniques: photometric --- planets and satellites: atmospheres  --- planets and satellites: terrestrial planets}


%
%
\section{Introduction}\label{sec:intro}
Characterizing the atmospheres on terrestrial exoplanets is one of the opportunities presented by near-future telescopes such as the \textit{James Webb Space Telescope} (JWST) \citep[e.g., ][]{morley2017observing} {and by potential future telescopes such as the \textit{Origins Space Telescope} (OST) that will have a significantly higher sensitivity in the mid-infrared than JWST}. Prior to observations, new theories on planetary atmospheres and a variety of climate modeling have been developed for understanding the potential atmospheres and interpreting future observations. For example, 
\citet{koll_temperature_2016} developed a radiative-convective-subsiding model that explains the temperature structure and atmospheric circulation on dry and tidally-locked rocky exoplanets. For moist atmospheres, most work focused on Earth-like thick atmospheres with 1 bar of N\2 and condensible water vapor in the habitable zone around M-dwarfs. 
\citet{yang_stabilizing_2013} first used a full three-dimensional global climate model (GCM) to study the H\2O runaway greenhouse on tidally-locked planets, and then \citet{yang_low-order_2014} constructed an idealized two column model to reproduce their GCM results. 

One interesting phenomenon in  \citet{yang_stabilizing_2013} is  when a tidally-locked habitable planet approaches the runaway greenhouse state, the night hemisphere emits more thermal radiation than the dayside seen from the broadband thermal phase curve. Such a reversal of thermal emission between dayside and nightside is confirmed by other GCM simulations \citep[e.g., ][]{wolf2017assessing, haqq2018demarcating}
and is unlikely to be produced by a dry atmosphere on tidally locked terrestrial planets \citep{koll_deciphering_2015}. If detected on a potentially habitable planet, it indicates two important features of the planetary atmosphere: (1) the emergence of high-level cirrus clouds associated with deep convection; (2) the cold trapping mechanism of water vapor is weakened so that optically opaque water vapor is allowed to build up in the night hemisphere and the nightside atmosphere emits thermal radiation from the warm mid-troposphere rather than from the cold surface \citep{yang_low-order_2014}. The former one confirms the presence of water vapor and its phase transition in the atmosphere, while the latter one precludes water vapor as a minor constituent in the atmosphere. 

The simulation described above in \citet{yang_stabilizing_2013} is only designed for a specific kind of atmosphere that is warm and thick with surface temperature ($T_s$) of 320\,K and background N\2 partial pressure of 1\,bar. The corresponding surface saturation mass concentration is  $\sim 10 \%$. The cold trap could still be effective if it is cold enough \citep{leconte_increased_2013}. Another reason for a reduced cold trap in \citet{yang_stabilizing_2013} is the warm air  at the cold trap ($\sim 270$\,K) due to the strong near-infrared absorption by water vapor. Then the mass concentration of water vapor at the cold trap reaches 3\%,  which is high enough to transport large amount of water vapor to the nightside. 
Other than raising the surface temperature or the cold trap temperature as mentioned above, the cold trapping mechanism can be greatly weakened by just removing the non-condensible constituent (e.g., N\2) from the atmosphere, even when both the surface and the cold trap are very cold \citep{wordsworth_water_2013}. Especially, a low N\2 atmosphere could occur on an M-dwarf rocky planet which lost most of its N\2 inventory early in its history, but could not generate a secondary N\2 atmosphere, because nitrogen is not easy to sequester in mantle minerals {while water can be well preserved in the mantle minerals due to its high solubility in the liquid magma ocean and be outgassed to the atmosphere by volcanism after the solidification of the magma ocean \footnote{CO\2 would also be outgassed along with H\2O. One issue is that it is unclear whether there are any circumstances in which a low N\2 atmosphere would create enough silicate weathering to keep down the CO\2 and maintain such a thin atmosphere. } (e.g., \citealt{moore2020water, kite2020secondary})}.

In this Letter, we use an idealized 3D GCM to simulate a thin and temperate condensible-rich atmosphere with an ineffective cold trap on slowly and synchronously rotating planets. We discuss the climatic similarity between the thin and temperate atmosphere and the warm and thick one. {The chief object of these simulations is to improve the understanding of the effect of cold-trapping and water vapor diluteness on the atmospheric circulation of tidally locked planets in the slowly-rotating regime, but the results have some interesting, if somewhat speculative because of uncertainty of cloud effects, implications for interpretation of thermal phase curves in terms of atmospheric composition.}

\section{Method}

The simulations are carried out by using the Exo-FMS general circulation model with simplified physical parameterizations including an active hydrological cycle based on formulations that remain valid regardless of the diluteness of the condensible substances in the atmosphere \citep{ding_convection_2016, pierrehumbert_dynamics_2016}. Our simulation design is the same as the study of condensible-rich atmospheres in \citet{pierrehumbert_dynamics_2016}, except that the planet is now synchronously rotating with the orbital period of 40 Earth days {so that the climate is in the slowly-rotating regime as discussed in \citet{haqq2018demarcating} and \citet{pierrehumbert2019tide}}. 

To weaken the cold trap of water vapor, the surface partial pressure of the non-condensible N\2 (hereafter denoted as $p_{a0}$) is chosen to be {0.03\,bar}, which is comparable to the saturation water vapor pressure corresponding to the substellar surface air temperature ($\sim$300\,K). To reach similar surface H\2O mixing ratio in thicker atmospheres, the corresponding $T_s$ is 373\,K if $p_{a0} = 1\,$bar; and 453\,K if $p_{a0} = 10\,$bar.  To investigate how the cold trapping of water vapor affects the thermal emission contrast between the day and night hemisphere, another two simulations with $p_{a0} = 1$ and 10\,bar are carried out. In all of the three simulations, the  fixed instellation {of 1200\,W\,m$^{-2}$} and background longwave optical opacity {of 1.2} are employed to maintain similar surface temperature distribution. The results we present are averages over the last 600 days of 3000-day integrations.

\section{Dynamic effects on water vapor distribution} \label{sec:dynamics}

\begin{figure*}[h]
  \centering
  \vspace{-10pt}
  \includegraphics[width=\textwidth]{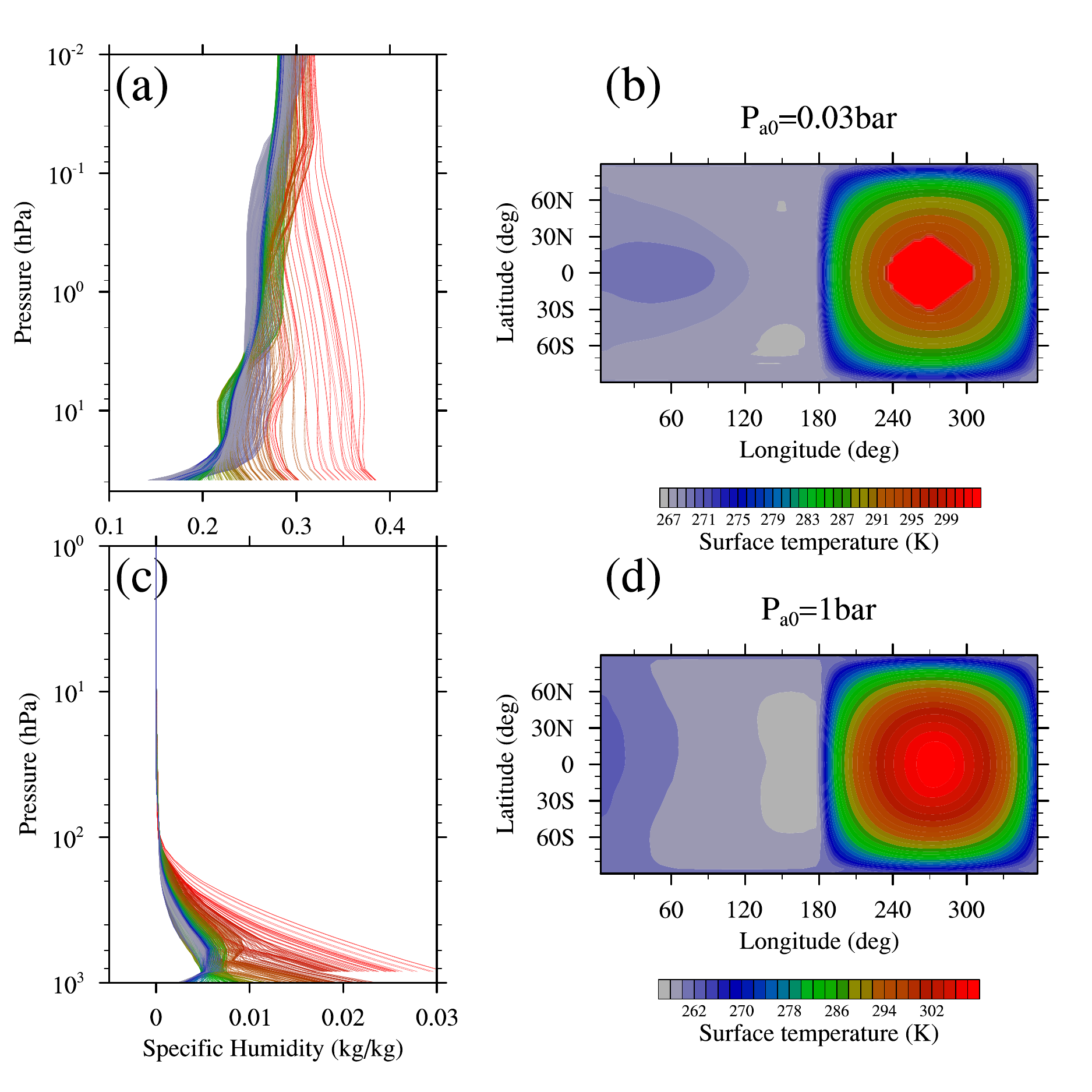}
  \caption{Upper panel: (a) the vertical profiles of the mass concentration of water vapor in the atmosphere and (b) the surface temperature distribution when \pa=0.03\,bar. Lower panel: same as the upper panel, but for \pa=1\,bar. The color of each profile in (a) and (c) is that of the corresponding surface temperature shown in (b) and (d), respectively. So red curves roughly represent the dayside deep convective region, while blue and gray curves represent the large-scale subsidence region.}\label{fig:vert}
\end{figure*}

Fig.~\ref{fig:vert}a shows the vertical distribution of water vapor mass concentration (hereafter denoted as $q$) in the non-dilute simulation when \pa = 0.03\,bar over the planet including the dayside deep convective region and the nightside large-scale subsidence region. The vertical distribution shares many similarities with the GCM simulations of the thick and warm climate in \citet{yang_stabilizing_2013}. 

First, in the dayside deep-convective region, the red curves in Fig.~\ref{fig:vert}a mostly follow the corresponding moist adiabats.
Compared to the dilute simulation when \pa=1\,bar in  Fig.~\ref{fig:vert}c, the vertical variation of $q$  is much smaller. This weak vertical $q$ gradient shows the weaker vertical temperature gradient for a nondilute moist adiabat has drastically reduced the cold trapping.  This vertical variation in the thin atmosphere is similar to that simulated in the thick and warm atmosphere shown in the log-pressure coordinate $(\log(p/p_s))$, as a consequence of the nature of moist adiabats. For moist adiabats, the vertical slope of $q$ (i.e., $\upd q/\upd \ln p$) is a decreasing function of both $1/T$ and $q$, and thus has a weak temperature dependence, and a very strong moisture dependence \citep{pierrehumbert_principles_2010, ding_convection_2016}. Since the slope of $q$ does not directly depend on   the background non-condensible partial pressure, the vertical distributions of $q$ along moist adiabats with the same surface $q$ are similar no matter the background atmosphere is thick or thin.

\begin{figure*}[h]
  \centering
  \vspace{-10pt}
  \includegraphics[width=\textwidth,trim=0 3cm 0 3cm,clip]{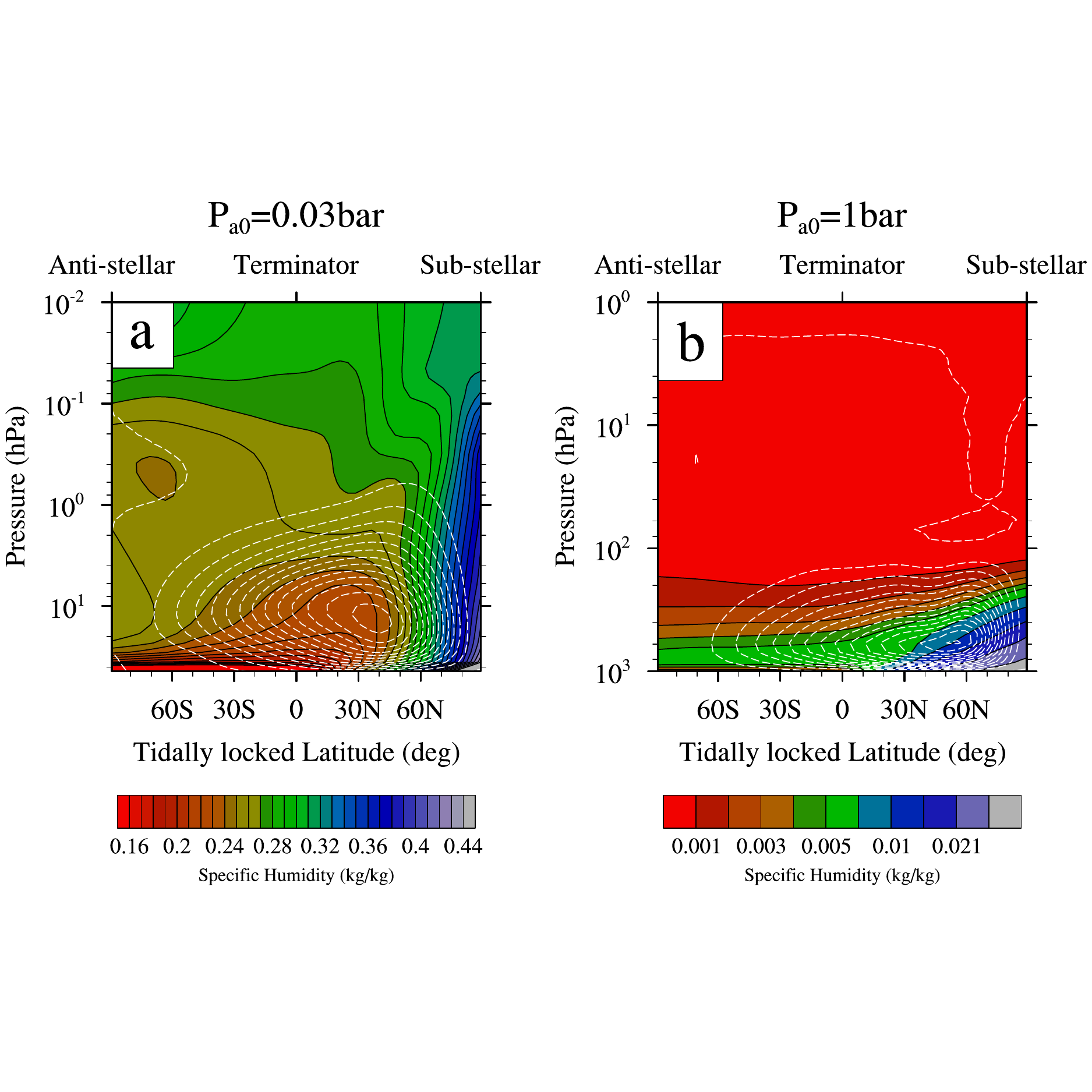}
  \caption{(a) Zonal-mean cross section of the specific humidity (color filled contours) and the meridional mass streamfunction (white dashed contours) in the tidally locked coordinate when \pa=0.03\,bar. (b) Same as (a), but for \pa=1\,bar. In both panels, a time-mean and zonal-mean counterclockwise circulation rising from the sub-stellar point is illustrated by the streamfunction.}\label{fig:stream}
\end{figure*}

Second, as the cold trapping of water vapor is weakened , the nightside atmosphere  becomes water-rich. To better understand this distribution,  the large-scale atmospheric circulation should be investigated. For slowly-rotating tidally-locked terrestrial planets, the atmospheric circulation is primarily dominated by a global thermally-direct overturning circulation that rises from the hot substellar region (white dashed contours in Fig.~\ref{fig:stream}), and superposed by perturbations due to  planetary waves 
\citep{merlis_atmospheric_2010, koll_temperature_2016}, except for the condensible-dominated atmosphere where sublimation/evaporation driven flow dominates \citep{ingersoll_supersonic_1985, castan_atmospheres_2011, ding2018pure}. In addition, the weak Coriolis force leads to quite horizontally uniform temperatures in the free atmosphere, which is referred to as the weak-temperature-gradient (WTG) approximation \citep{pierrehumbert_dynamics_2016}. For the run with thinnest atmosphere  (\pa=0.03\,bar), WTG still roughly applies and the hemisphere-averaged air temperature difference is less than 5\,K in the free troposphere between day and night side (not shown).   
As a result of the overturning circulation, most of the air in the night hemisphere has to pass through the substellar tropopause -- the driest place in the free atmosphere due to the WTG approximation. So the substellar tropopause behaves as a cold trap of water vapor. After the air leaves the cold trap, condensation never occurs until it enters the strong near surface temperature inversion layer on the nightside. In this case, water vapor is no different from a passive tracer, and its concentration in an air parcel is conserved along any segment of a trajectory assuming weak mixing with neighboring parcels. This concept of advection-condensation model was used to reconstruct the relative humidity in the Earth's subtropics by large-scale wind and temperature fields, and showed excellent agreement with satellite observations \citep{pierrehumbert_evidence_1998}. The model was also used to explain the simulation results of the relative humidity distributions from a variety of GCMs in \citet{yang2019intercomparison}.
Similar reasoning explains the vertical $q$ distribution on the nightside as the gray curves in Fig.~\ref{fig:vert}a shows: $q$ roughly remains uniform between the top of the model and 10\,hPa; as the air continues to subside and reaches the low-level temperature inversion layer, condensation occurs again and $q$ starts to decrease resulting in the relatively dry region shown as the red area in Fig.~\ref{fig:stream}a. 

The advection-condensation model is also roughly illustrated in Fig.~\ref{fig:stream} by overlaying the zonal-mean cross-section of $q$ with the meridional mass streamfunction in the tidally-locked coordinate (see Appendix B in \citet{koll_deciphering_2015} for details about the tidally-locked coordinate). Within 30$^\circ$ around the substellar point, the contours of $q$ intersect with the streamlines due to condensation. As the air parcel leaves the upwelling branch of the overturning circulation, the contours of $q$ start to coincide with the streamlines. Note that the results in Fig.~\ref{fig:stream} is for the time and zonal mean, so in some places on the nightside the contours of $q$ still intersect with the streamlines where the planetary wave perturbations are strong compared to the mean flow and the mean overturning circulation is no longer a good representation of the true Lagrangian trajectories. 
In fact, our simulation of the non-dilute atmosphere in Fig.~\ref{fig:stream}a shows many interesting dynamic features other than a simple overturning circulation. The large-scale planetary waves are highly variable in time and are likely  convectively-coupled inertial-gravity waves, which makes it difficult to characterize the statistics of air parcel trajectories. In addition, upwelling motion associated with inertia-gravity waves occurs in the lower atmosphere on the nightside  at times, which further cools and dehydrates the atmosphere. In this case, the nightside low-level temperature inversion layer serves as another cold trap of water vapor. Detailed discussion of the dynamics of the non-dilute atmosphere is beyond the scope of this Letter. 

\section{Broadband thermal phase curves} \label{sec:phase}

\begin{figure}[ht]
  \centering
  \vspace{-10pt}
  \includegraphics[width=0.6\columnwidth]{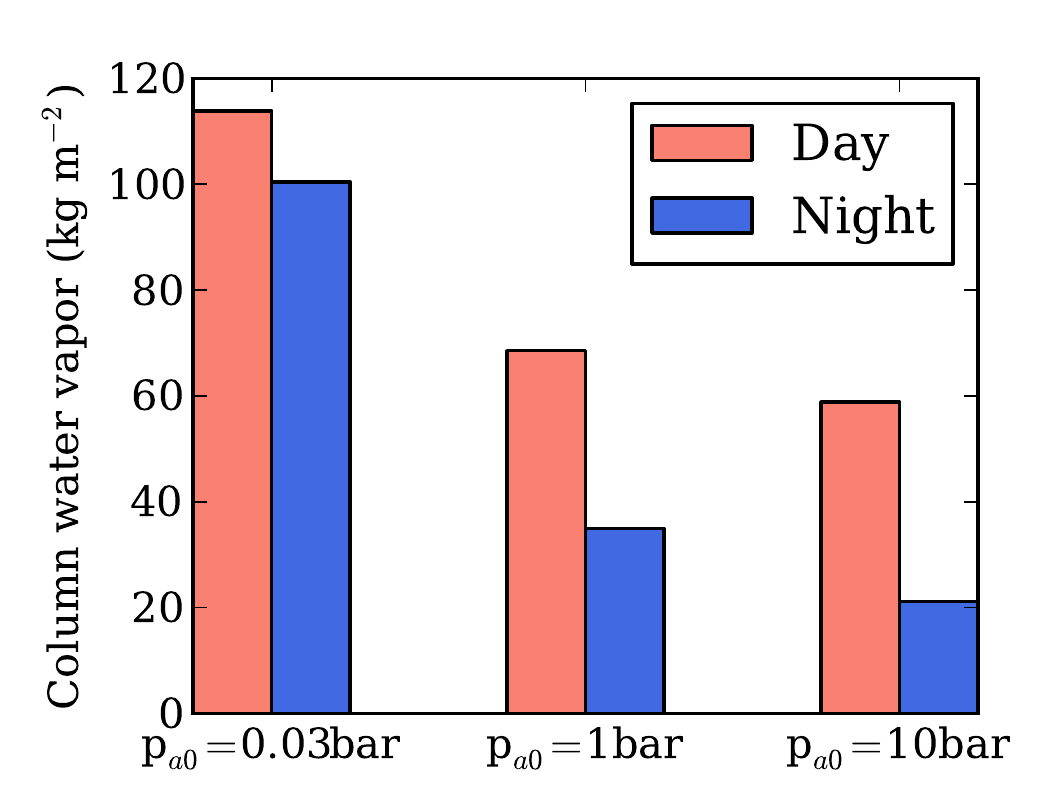}
  \caption{The vertically integrated mass of water vapor averaged over the day and nightside hemisphere, when \pa = 0.03, 1, 10\,bar, respectively. }\label{fig:column}
\end{figure}

To investigate the observational consequence of the climate with a reduced cold trap, we focus on the broadband thermal phase curve around 10\,$\mathrm{\upmu m}$ where most energy of the terrestrial radiation locates. In this spectral region, the atmospheric absorption is dominated by the water vapor self-induced continuum. Again, like the slope of the moist adiabat, self-induced continuum depends little on  the background air pressure, but  only on the vertical  integration of water vapor mass \footnote{Strictly speaking, the opacity $\tau \propto \int \rho_v^2 \upd z$, rather than the  vertical  integration of water vapor mass $\int \rho_v \upd z$}.

Fig.~\ref{fig:column} shows the column-integrated mass of water vapor averaged over the day and night hemisphere, respectively. First, the contrast of column water vapor mass between the day and night hemisphere decreases as the atmosphere is thinner and the cold trap becomes less effective. This trend is linked to the change of the vertical slope of the moist adiabat discussed in Section~\ref{sec:dynamics}, which has direct impact on the contrast of thermal emission between the day and night hemisphere, and will be discussed later. Next, the column water vapor mass on the dayside
increases as the atmosphere is thinner. It may seem contour-intuitive, but in fact is also related with the fact that water vapor is a non-dilute component in the atmosphere. To show it in a simple way, assuming that the entire atmospheric profile is governed by the moist adiabat, then the column water vapor mass can be approximately written as 
\begin{equation}
m_v \approx \int_{0}^{p_s} q(p)\ \upd p /g = \int_{0}^{T_s} f(r, 1/T) \ \upd T /g, \qquad r = p_v/ p_a
\end{equation}
where $g$ is surface gravity, $q$ and $r$ are the mass concentration and molar mixing ratio of water vapor, $p_v$ and $p_a$ are partial pressures of water vapor and the background air, and $f(r, 1/T)$ is an increasing function of both $r$ and $1/T$. The  temperature range of the  three runs in Figure~\ref{fig:column} are similar, but molar mixing ratio varies over several orders of magnitude, which makes the thinner atmosphere in fact contain more water vapor. 

\begin{figure}[h]
  \centering
  \vspace{-10pt}
  \includegraphics[width=\columnwidth]{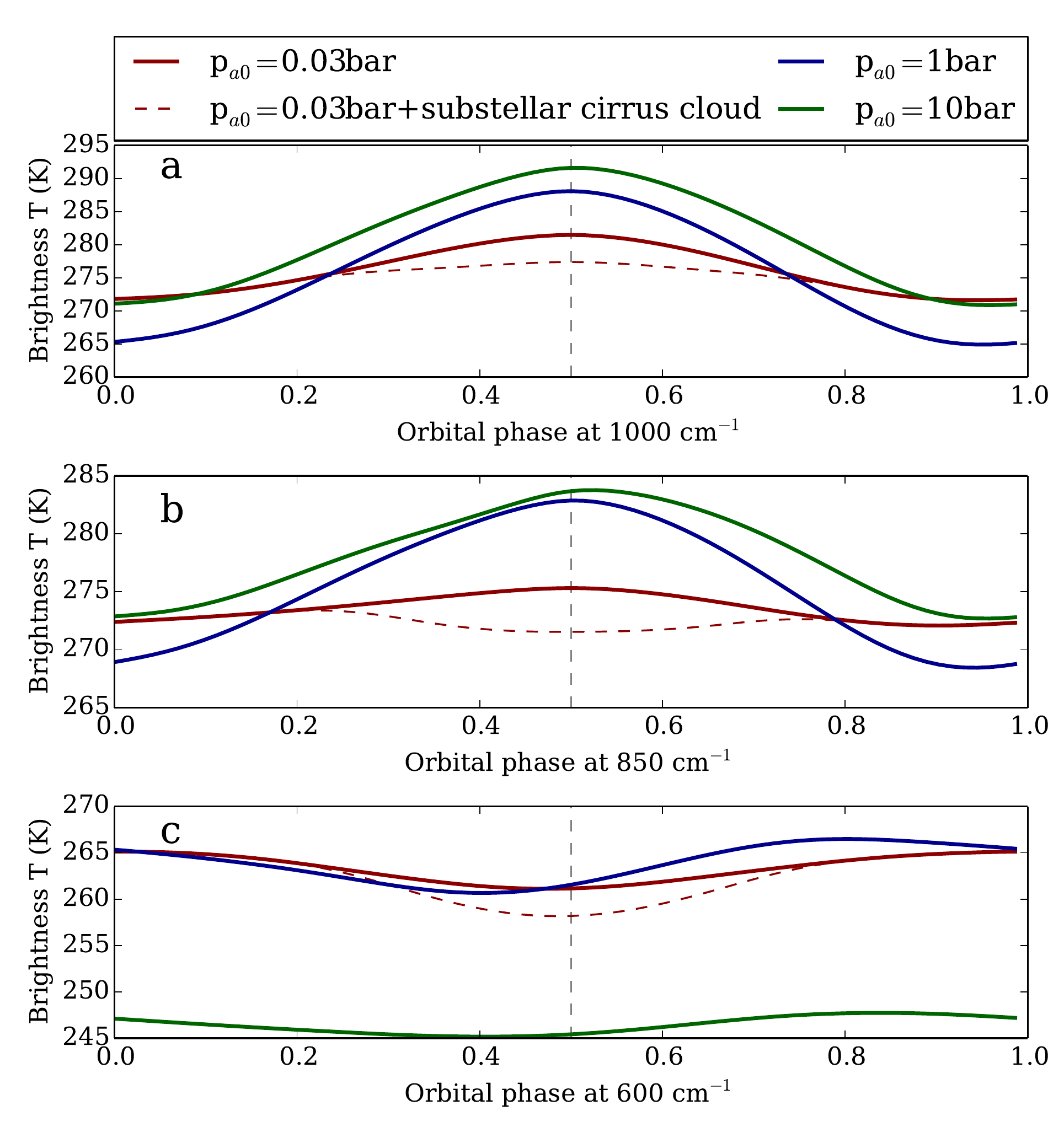}
  \caption{Broadband thermal phase curve expressed as the brightness air temperature at three spectral bands when \pa=0.03\,bar (red solid), 1\,bar (blue solid), 10\,bar (green solid), respectively: (a) at 1000\,$\mathrm{cm^{-1}}$; (b) at 850\,$\mathrm{cm^{-1}}$; (c) at 600\,$\mathrm{cm^{-1}}$. The bandwidth of each band is 50\,$\mathrm{cm^{-1}}$. The red dashed curves are the thermal phase curves when \pa=0.03\,bar and a cirrus cloud layer was prescribed near the tropopause within 30$^\circ$ around the substellar point with an optical thickness of 0.5.
  The gray dashed line in each panel marks the superior conjunction.}\label{fig:bt}
\end{figure}

 Exo-FMS uses a two-stream gray gas radiative scheme because it is computationally cheap and meanwhile the simulation result is similar to that with realistic radiative schemes. The model is also cloud-free since we focus on the most basic dynamical implications without confronting all the issues of cloud microphysics in nondilute atmospheres. However, the cloud information   can be diagnosed by the condensation profile of water vapor in our GCM simulations. For example, the deep convection within 30$^\circ$ around the substellar point in the simulation when \pa=0.03\,bar indicates the formation of high-level cirrus clouds there, and the near surface condensation in the night hemisphere indicates the formation of low-level stratus clouds. In this section, we will first investigate the radiative property of the weak cold-trap climate with a realistic clear-sky radiative model and then take into account the radiative effect of clouds. This neglects the feedback of clouds on the circulation, but it is commonly done in {hot gas giant} exoplanet studies, such as the work on hot Jupiter clouds by \citet{parmentier2016transitions}.

We calculate the broadband thermal phase curve in three bands which center at 1000, 850 and 600\,cm$^{-1}$ with bandwidth of 50\,cm$^{-1}$. In these spectral range, the water vapor self-continuum absorption is used by the polynomial fit given by \citet{pierrehumbert_principles_2010}. 
We first compute the disk-averaged radiative flux based on the 3D simulated temperature and humidity fields, and then translate it into the brightness temperature, as shown in Fig.~\ref{fig:bt}. For water vapor self-continuum, the equivalent path is quadratic in water vapor partial pressure, which makes the contrast of thermal emission between the day and night hemisphere sensitive to the contrast of column-integrated mass of water vapor shown in Fig.~\ref{fig:column}. For the weak cold-trap climate with \pa = 0.03\,bar, optically-opaque water vapor atmosphere can build up in both day and night hemispheres associated with the small contrast of column-integrated water vapor mass, resulting in similar emission levels between the day and night hemisphere and the small brightness temperature variations of the thermal phase curve in all three bands. On the contrary, in the two dilute runs with effective cold traps, the optically-thin atmosphere can only emit infrared radiation to space from the surface or the low-level atmosphere 
unless in very strong absorption band. In fact, the water vapor absorption at 600\,cm$^{-1}$ is so strong that the nightside is optically opaque within this band in  all the three runs. As a result, Fig.~\ref{fig:bt}c shows that all three runs have small brightness temperature variations with a small degree of reversal in thermal emission between the day and night hemisphere. 

If the radiative effect of clouds is taken into account in the model, the thermal phase curve integrated over all frequencies in the weak cold-trap climate with a thin and temperate atmosphere should show similar feature as the one of the thick and warm atmosphere simulated by \citet{yang_stabilizing_2013} because the cloud distributions in the two climates with weak cold-trap are similar. The near surface stratus clouds on the nightside would contribute little to the thermal emission because the emission of water vapor is roughly at the same level, while the high-level convective cirrus clouds on the dayside would reduce the thermal emission so that  a reversal between day and night hemisphere is expected.  Nevertheless, future work by using full 3D GCM with realistic radiative transfer calculation (and cloud-resolving grids if possible) is required to validate it. 

Without simulating the complex cloud behavior in full 3D GCM, we can estimate its impact by calculating the phase curves with our clear-sky GCM run with \pa = 0.03\,bar and a prescribed substellar cirrus cloud layer. \added{The cloud layer is within 30$^\circ$ around the substellar point and between 10 and 20 hPa, which is the condensation region near the substellar tropopause as shown in Fig.~\ref{fig:stream}a.} The red dashed curves in Fig.~\ref{fig:bt} show that the brightness temperature can be reduced by $\sim$5K at the superior conjunction, leading to the reversal of the thermal phase curve, when the cirrus cloud optical thickness is 0.5 (a typical value of the cirrus cloud  in the present Earth's tropical climate). Moreover, if the amplitude of the thermal phase curve was 5K, the equivalent band-integrated (10-28 $\mu$m) contrast for the Mid-Infrared Instrument (MIRI) on
{JWST} would be 4 ppm following the assumptions in \citet{yang_stabilizing_2013}  \added{for M-dwarf habitable zone terrestrial planets 5pc away}. \replaced{Note that the expected one-day photometric precision for transiting M-Dwarf habitable zone terrestrial planets is 1 ppm in \citet{yang_stabilizing_2013}. So the reversal of the thermal phase curve could be potentially detectable by {JWST}.} {The expected photometric precision of a one-day integration with JWST  is 0.97ppm following  \citet{yang_stabilizing_2013} and corrections in \citet{koll_deciphering_2015}, but the expected noise floor for MIRI is 50ppm \citep{greene2016jwst, suissa2020dim}.  So the reversal of thermal phase curve is unlikely to be detected by MIRI if MIRI has such a significant noise floor, while it could be potentially detectable by  OST because OST will baseline a noise floor of 5ppm \citep{meixner2019ost}.  }

\section{Conclusion and Discussion} \label{sec:conclusions}

Compared to the simulations of the thick and warm atmosphere on slowly and synchronously rotating planet by \citet{yang_stabilizing_2013}, our simulation of a thin and temperate atmosphere with a weak cold trap 
also shows a climate that is dominated by a thermally-direct overturning circulation with formation of substellar clouds and build up of optically opaque water vapor on the nightside. Therefore we propose if the signal of the reversal in the broadband thermal phase curve is detected on potentially habitable exoplanets, it is very likely that the atmosphere have a weak cold trap which could be either the thick and warm atmosphere as in \citet{yang_stabilizing_2013} or the thin and temperate atmosphere in our study. This signal provides  information regarding the relative ratio of water vapor in the atmosphere, unlike the transit spectroscopy that measures the absolute quantity of water vapor. 
It is equivalent to say that the reversal in the broadband thermal phase curve implies the total atmospheric pressure should be below a certain threshold relative to the saturation vapor pressure at the surface to weaken the cold trap.  {This adds new observational constraint on the atmospheric pressure near the substellar tropopause or the emission level on the nightside if the air temperatures there could be detected by thermal emissions. Particularly, in our simulation on the thin and temperate atmosphere, the nightside emission level in the window region of water vapor is not far above the ground. This threshold of water vapor concentration needs to be explored by a full 3D GCM with real-gas radiative transfer  and interactive cloud calculations, and cannot be identified accurately by our idealized GCM simulations.}

{Next}, our simulation of the thin condensible-rich atmosphere is not exactly dynamically analogous to the thick and warm case in  \citet{yang_stabilizing_2013} in terms of the nondiluteness of moisture. There is stronger nondilute effect of water vapor in our simulation, and the contribution of water vapor on the air density and the atmospheric pressure becomes more important. One major consequence of the nondilute effect is if we continue to decrease the non-condensible inventory in the atmosphere to certain extent, the overturning circulation may not be maintained anymore because it requires the surface pressure gradient force pointing towards the day hemisphere against the surface friction. In this case, the circulation should reduce to the sublimation/evaporation-driven flow that is dominated by low-level atmospheric transport from the hot dayside to the cold nightside \citep{ingersoll_supersonic_1985, castan_atmospheres_2011, ding2018pure}. {Conventional 3D GCMs, including Exo-FMS, cannot handle such strong contrasts in surface pressure, due to numerical issues arising from the vertical coordinate system. }

Finally, planetary rotation rate plays a fundamental role in determining the morphology of thermal phase curves. For M-dwarfs with effective temperature of less than 3000 K, the rotation period of habitable planets is typically  $\sim$ 5 Earth days \citep{haqq2018demarcating}. The fast spin rate of such rapid rotators makes their Rossby deformation radii  less than their planetary radii, so perturbations due to stationary planetary waves will alter the circulation pattern \citep{wolf2017assessing, haqq2018demarcating} and our discussion based on WTG and cold trapping of moisture in an overturning circulation will not apply. For example, \citet{haqq2018demarcating} showed that on a synchronously rotating exoplanet with rotation period of 4.25 days the cirrus cloud deck is shifted towards the eastern  terminator by the advection of super-rotating equatorial jet (Fig.~11 in \citealt{haqq2018demarcating}) and the dayside still emits more infrared radiation to space than the nightside. However, as in \citet{yang_stabilizing_2013}, \citet{haqq2018demarcating} also assumed a background N\2 partial pressure of 1 bar. Whether similar cloud behavior would occur in a thin but water-rich atmosphere on a rapidly rotating planet is an interesting problem to be addressed in future work.

\acknowledgments
 
We thank the referee for thoughtful comments that improved the manuscript. Support for this work was provided by the NASA Astrobiology Institutes Virtual Planetary Laboratory Lead Team, under the National Aeronautics and Space Administration solicitation NNH12ZDA002C and Cooperative Agreement Number NNA13AA93A, and by the European Research Council Advanced Grant EXOCONDENSE (Grant 740963). 
The numerical calculation was completed in part with resources provided by the University of Chicago Research Computing Center. 
 

%

\bibliography{non_dilute}

\bibliographystyle{aasjournal}
\listofchanges


\end{document}